\documentstyle[twocolumn,aps]{revtex}
\input epsf
\begin{document}

\twocolumn[\hsize\textwidth\columnwidth\hsize\csname@twocolumnfalse\endcsname
\title{Experimental Investigation of Resonant Activation}

\author{Rosario N. Mantegna and Bernardo Spagnolo}

\address{Istituto Nazionale per la Fisica della Materia, 
Unit\`a di Palermo\\ and\\
Dipartimento di Energetica ed Applicazioni di Fisica,
Universit\`a di Palermo, Viale delle Scienze, I-90128,
Palermo, Italia}


\maketitle

\begin{abstract}

We experimentally investigate the escape from a metastable state over a 
fluctuating barrier of a physical system. The system is switching between 
two states under electronic control of a dichotomous noise. We measure 
the escape time and its probability density function 
as a function of the correlation rate of the dichotomous noise in a 
frequency interval spanning more than 4 frequency decades. We observe 
resonant activation, namely a minimum of the average escape time as a 
function of the correlation rate. We detect two regimes 
in the study of the shape of the escape time probability distribution: 
(i) a regime of exponential and (ii) a regime of non-exponential 
probability distribution.

\end{abstract}


\vskip1pc]

The thermal activated escape of a particle over a potential barrier in a 
metastable state is a classical problem since the seminal work of Kramers 
\cite{Kramers40}. It occurs in a wide variety of physical, biological and 
chemical systems \cite{Hanggi90}. When the height of the potential barrier 
is itself a random variable the dynamics of the escape from the metastable
state is affected by the statistical properties of the random variable 
controlling the barrier height. In the presence of barrier fluctuations 
rather counterintuitive phenomena may occur. A classical example is the 
phenomenon of resonant activation \cite{Doering92}.
The signature of the resonant activation phenomenon is observed when the 
average escape time of a particle from the metastable state exhibits a 
minimum as a function of the parameters of the barrier fluctuations.
The original work of Doering and Gadoua triggered a large amount of 
theoretical activity 
\cite{Zurcher93,Broeck93,Bier93,Brey94,Pechukas94,Madureira95,Reimann95,Iwariszewski96,Reimann97,Boguna98,Reimann98,Mantegna98}. Analogic simulations of resonant 
activation have also been performed in a bistable system in the presence 
of multiplicative Gaussian or dichotomous noise 
\cite{Marchesoni95,Marchi96}.

This letter aims to answer the following questions - (i) is resonant 
activation experimentally observable in a physical system? (ii) what 
about the probability density function of escape time?
We answer both questions by investigating the escape from a metastable 
state over a fluctuating barrier in a physical system. The physical system 
is a tunnel diode biased in a strongly asymmetric bistable state in the 
presence of two independent sources of electronic noise. Specifically 
we have a dichotomous noise source controlling the metastable potential 
and a Gaussian noise source mimicking the role of a thermal source.
In this physical system we observe resonant activation and we measure 
and characterize the probability density function (pdf) of escape time
by finding two distinct regimes. 

Our physical system is the series of a resistive network $R$ with a tunnel 
diode in parallel to a capacitor (in our case the sum of diode capacitance 
and input capacitor of the measuring instrument). 
This system is rather versatile and allows investigate the dynamics 
of a bistable or metastable system in the presence of noise. It has been 
used to investigate stochastic resonance \cite{Mantegna94} and 
noise enhanced stability \cite{Mantegna96}.

The equation of motion of such system is
\begin{equation}
\frac{d v_d}{dt}=-\frac{d U(v_d)}{dv_d} +\frac{1}{RC} v_n(t)
\end{equation}
with
\begin{equation}
U(v_d)=-\frac{V_B v_d}{RC}+\frac{v_d^2}{2RC}+\frac{1}{C}\int_0^{v_d} I(v) dv ,
\end{equation}
where $R$ is the biasing resistor of the network, $C$ is the 
capacitance in parallel to the diode, $V_{rms}$ is the
amplitude of the Gaussian noise $v_n(t)$, $V_B$ is the biasing 
voltage and $I(v)$ is the nonlinear current-voltage 
characteristic of the tunnel diode. 
The effective associated potential is a bistable potential.
We control the parameters of our electronic network
($V_B$ and $R$) in such a way that one of the two wells is 
much deeper than the other. By setting such strong asymmetry between 
the two wells we essentially deal with a metastable state. In fact, 
the probability that the system once escaped goes back into the 
metastable state during the experimental time is negligible.
By using a digital electronic 
switch we vary the value of the series resistance of our metastable 
system between the two values $R_+$ and $R_-$. When the system 
switches from $R_+$ to $R_-$ two effects take place. One concerns 
the variation of the height of the barrier of the metastable 
state whereas the other consists in a slight change of the value of 
intrinsic time scale of the system (the $RC$ 
term of Eqs. (1) and (2)). In our experiment we use a germanium tunnel 
diode 1N3149A, which has nominal peak current $I_p=10.0$ $mA$, 
peak voltage $V_p=60$ $mV$, valley current $I_v=1.3$ $mA$, 
valley voltage $V_v=350$ $mV$ and a rather short switching time 
(less than $50$ $nsec$). We perform our experiments by choosing 
$V_b=8.95$ $V$, $R_+=1100$ $\Omega$, $R_-=1080$ $\Omega$ and $C=45$ $pF$. 
The noise $v_n(t)$ is obtained 
starting from a digital pseudo-random generator. The noise 
voltage is a stochastic Gaussian noise synthesized by a 
commercial source (the noise generator DS345 of Stanford 
Instruments). The Gaussian noise $v_n(t)$ is added to the 
bias signal $V_B$ by an electronic adder based on a low-noise
wide band operational amplifier. At the output of the 
operational amplifier we measure the noise $v_n(t)$. It is
a Gaussian noise characterized by a spectral density which is
flat at low frequency ($f<1$ MHz), having a moderate 
(approximately 7 dB) increase in the region ($1<f<3$ MHz) and 
quickly decreasing after the cut-off frequency ($f_{cut-off}$=4.6 MHz).
By defining the correlation time of the Gaussian noise 
as the time at which the normalized autocorrelation function 
assumes the value $1/e$, we measure $\tau_n=68$ ns.
In addition to the Gaussian noise source, a 
dichotomous noise source is also present.
The dichotomous noise source is obtained by using the commercial 
chip MM5437 of National semiconductor. An external clock drives 
the dichotomous noise source and allows us to 
control the noise correlation time over a wide range. 
The normalized autocorrelation function of this pseudorandom 
noise is linearly decreasing from 1 to zero in one clock 
period $1/f_c$. After this time the autocorrelation function 
is equal to zero within the experimental errors. 
By defining the correlation time as before,
we have that $\gamma \equiv \tau_C^{-1} \simeq 1.3 \times f_c$. 
In our experiments the correlation rate $\gamma$ is varied 
in the range from $\gamma_{min}=18$ Hz to $\gamma_{max}=316228$ Hz. This is an
experimental interval consisting of more than 4 frequency decades.
To investigate a so wide range of frequency we have to
overcome two experimental conflicting constraints:
(i) we are forced to set the time constant of our system 
$\tau_s \equiv RC$ to a low value satisfying the 
inequality $\gamma_{max} \ll 1/ \tau_s$ and (ii) we need to 
use a high value of $\tau_s$ to maintain the ratio 
$\tau_n/\tau_s$ as low as possible to conduct our experiments
in the``white noise" limit of $v_n(t)$. The best compromise
we find is to set $\tau_s \le R_+ C=49.5$ ns. With this choice 
$1/ \tau_s \approx 50 \gamma_{max}$ and $\tau_n/\tau_s \approx 1.40$.
In other words we guarantee the investigation of the resonant
activation phenomenon in a wide range of frequency by performing
our experiments in a regime of moderately colored noise. 

Under computer control, we measure, for each value of the 
correlation time of the binary noise, $5 \times 10^3$ escape 
times from the metastable fluctuating state. In each statistical 
realization, the system is put into the metastable state 
by an electronic switch. We define the escape time $T$ as 
the time interval measured between the setting of the system 
in the metastable state ($t=0$) and the crossing of $v_d$ of 
a voltage threshold. The selected voltage threshold 
($v_d=0.4$ $V$) ensures that the system is quite far from the 
starting well. The exact value of the threshold is not a 
determinant parameter because the escape from the well is fast
(of the order of $\tau_s$). From the set of measured escape 
times $T$, we determine the average value $<T>$ and 
the distribution $P(T)$. The first measurement is devoted 
to determine the Kramers rates 
as a function of the noise amplitude $V_{rms}$ for the two 
metastable states in the {\it absence} of the dichotomous noise 
modulation. In our experiments we vary $V_{rms}$ in the 
interval from 0.816 V to 1.00 V and we
observe that the logarithm 
of the average escape time is a linear function of the 
inverse of the 
noise intensity $1/V^2_{rms}$ for the two states 
``+" and ``-" of 
the metastable state. 
In other words, the measured values lay in a straight line 
verifying the Kramers law in both cases. The two metastable 
states are characterized by 
different slopes of the two lines fitting 
experimental data. These experimental observations 
confirm that the shape of the metastable potential 
is different in the two states ``+" and ``-". 
The observation of a Kramers law is in agreement with the 
theoretical prediction of the escape from a metastable state
both in the absence and in the presence of colored noise.
Specifically, in a regime of colored 
noise $v_n(t)$, the expected functional form is still 
Kramers-like \cite{Bray89,Dykman90} and can be written as 
$<T>=C~\exp{[\Delta U_{m} \tau_s^2/V^2_{rms}\tau_n]}$
\cite{Hanggi85}. 
Here $C$ is a prefactor and $\Delta U_{m}$ is the 
measured potential barrier height associated to a system 
characterized by a time constant $\tau_s$ and evolving 
in the presence of a colored noise with correlation 
time $\tau_n$. Theoretical studies predict that 
$\Delta U_{m}$ is a function of $\tau_n/\tau_s$ 
\cite{Bray89,Dykman90,Hanggi85}.
We verify that we are in a regime of colored noise
by performing an experimental test \cite{note1}.

We perform our investigation of the average escape time as 
a function of the correlation rate of the dichotomous noise 
for two different values of the amplitude of the 
Gaussian noise. The chosen values are $V_{rms}=0.816$ V 
and $V_{rms}=0.852$ V. 
In the absence of switching between the two states, the 
average escape times from state ``+" and ``-" are 
$<T_+>=0.030$ s and $<T_->=0.0021$ s when $V_{rms}=0.816$ V 
and $<T_+>=0.0082$ s and $<T_->=0.00081$ s when 
$V_{rms}=0.852$ V. In the presence of switching between the 
two states the behavior is different. In Fig. 1 we show  
the average escape time as a function of $\gamma$ in a 
log-log plot. We also indicate the value of the Kramers 
rates $\mu_+=1/<T_+>$ and $\mu_-=1/<T_->$ for the two sets 
of experiments. In both cases 
the distinctive characteristics of resonant activation are 
observed \cite{Doering92}, the average escape time initially 
decreases, reaches a minimum value and then again increases 
as a function of $\gamma$. Specifically for values of 
the correlation rate $\gamma$ less 
than $\mu_+$ the average escape time approaches the value 
$<T>=(<T_->+<T_+>)/2$ predicted by resonant activation 
theory \cite{Doering92,Bier93,Boguna98}. We address this regime 
as the regime of slow dynamics of the barrier height. For values 
of the correlation rate approximately ten times higher 
than $\mu_-$, the minimum value of the average escape time is observed.
Here resonant activation is fully effective. In this regime, the 
value of the average escape time is in both cases 
approximately 10\% less than the one expected from resonant 
activation theories in the kinetic approximation 
\cite{Bier93,Boguna98}, namely $<T>=((\mu_+ +\mu_-)/2)^{-1}$. 
We have not a definitive explanation for this discrepancy. 
One possibility is that this effect reflects the fact that in 
our experimental set-up our system switches between 
two states in a regime of colored Gaussian noise rather 
than in the regime of ``white noise". A second one is that 
this behavior is related to the fact 
that our system switches between 
two states which in addition to a different potential barrier 
height are also characterized by different shape of the 
potential and different time constant.
For values of $\gamma$ higher than 10 kHz the average escape 
time starts to increase as a function of $\gamma$ and approaches 
the value associated with an effective potential characterized 
by an average barrier at the highest value of $\gamma$. This 
is the regime of fast dynamics of the barrier height. A comparison 
of the two sets of experiments of Fig. 1 also shows that the 
region of resonant activation becomes more flat when the noise 
intensity decreases. For our experiments the barrier heights 
are kept constant in the two series of experiments. This 
observation is complementary to the theoretical observation 
that the resonant activation region becomes more flat when 
the barrier height is increased while the noise intensity 
is kept constant \cite{Reimann97,Boguna98}.

For each value of the noise intensity $V^2_{rms}$ and for 
each value of the correlation rate we measure $P(T)$, the pdf of 
the escape time. In Fig. 2 we show all the pdfs measured 
with $V_{rms}=0.816$ V in a 3-dimensional semi-logarithmic 
plot. The regimes of slow dynamics, resonant activation and 
fast dynamics of the dichotomous noise are clearly observed 
moving from the left to the right.
Concerning the shape of the pdf, our results show that two 
distinct regimes are present. In particular for values of
$\gamma$ greater than $\mu_-$,  $P(T)$ is well described 
by an exponential decay function
$P(T)=\exp{[-T/<T>]/<T>}$ whereas in the 
opposite regime the pdf is non-exponential.
Hence, in the resonant activation regime, the pdf has an 
exponential shape.
This experimental observation is in agreement with the 
numerical observation that the system escapes preferentially 
through the state with the lowest barrier at the resonant 
activation \cite{Doering92}.
In other words, at the resonant activation the system 
approximately experiences a single potential barrier and 
this ends up into an exponential pdf. 
Exponential pdfs are also observed for higher values of 
$\gamma$ in the fast dynamics regime.
The experimental detection of an exponential pdf for high 
values of $\gamma$ is in agreement with the simple 
description that the system experiences an effective 
potential with an average barrier for the highest values 
of the correlation rate.

The final investigation of our study concerns the degree 
of non-exponential behavior observed when $\gamma<\mu_-$. 
To quantify the difference between the real pdf and an 
exponential pdf we  measure the standard deviation of 
the average escape time.
For an exponential pdf the two observables coincide, 
hence a difference in these observables manifests a 
non-exponential behavior of the pdf.
In Fig. 3 we show the average escape time together with 
its standard deviation as a function of $\gamma$ measured 
by setting $V_{rms}=0.816$ V . The two regimes are 
clearly seen. For high values of $\gamma$ the pdf is well 
described by an exponential function, here the average 
escape time and its standard deviation essentially coincide. 
When the correlation rate is less than $\mu_-$ a deviation 
from the exponential form starts to emerge and becomes more 
pronounced for lowest values of $\gamma$. In particular, 
for very low values of the correlation rate ($\gamma<\mu_+<\mu_-$) 
the pdf assume the form of a double exponential 
\begin{eqnarray}
P(T)=\frac{1}{2<T_+>} \exp{[-T/<T_+>]}+ \nonumber \\
\frac{1}{2<T_->} \exp{[-T/<T_->]} . 
\end{eqnarray}
The inset of Fig. 3 shows $P(T)$ measured when $\gamma=13$ Hz 
in a semi-logarithmic plot. The shape is clearly non-exponential. 
It is well approximated by Eq. (3) when the measured values of 
$<T_+>$ and $<T_->$ are used (solid line in the inset). 
In this regime the correlation rate is smaller than Kramers 
rates of both states. This implies that the system essentially 
starts and ends each realization of the stochastic process in 
one of the two states with probability 1/2 and the overall 
pdf is given by Eq. (3).

In this letter we present an experimental study detecting 
resonant activation and we verify some 
theoretical results obtained in model systems for limit 
regimes of the correlation rate. Resonant activation is 
observed in spite of the fact that the Gaussian noise mimicking
the presence of temperature is colored instead of being ``white".
In our opinion, our result shows that the resonant 
activation phenomenon is pretty robust and may be 
observed in a variety of physical systems. By 
investigating the pdf of the average escape time, we detect 
a non-exponential shape of the pdf in the regime of 
low values of $\gamma$. 
This finding suggests that systems characterized by 
a non-exponential escape time probability distribution could 
be simply modeled in terms of metastable system with a 
fluctuating barrier.

We wish to thank INFM, ASI (contract ARS-98-83) and MURST 
for financial support.

\begin{figure}
\epsfxsize=3.0in
\epsfbox{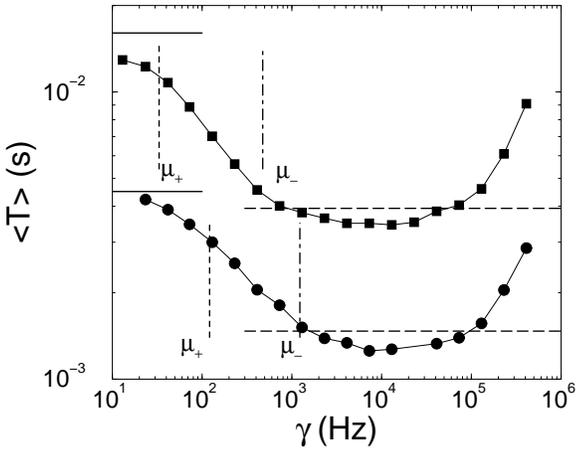}
\narrowtext
\caption{Average escape time as a function of the correlation 
rate of the dichotomous noise. The amplitude of the 
Gaussian noise is set to $V_{rms}=0.816$ V (squares) 
and $V_{rms}=0.852$ V (circles). For both sets of data, 
dashed and dot-dashed lines indicate the Kramers rate of 
the two states ``+" and ``-" respectively. Three regimes 
are clearly seen: (i) the regime of slow dynamics between 
states ($\gamma<\mu_+$); (ii) the regime of resonant 
activation ($10^3<\gamma<10^5$ Hz) and (iii) the regime 
of fast dynamics between states ($\gamma>10^5$ Hz). 
The solid lines indicate the asymptotic behavior expected 
for slow dynamics whereas the long-dashed lines indicate 
the asymptotic value expected from the kinetic approximation. 
The regime of resonant activation is broader for the 
value $V_{rms}=0.816$ V (squares).
\label{f.1}}
\end{figure}

\begin{figure}
\epsfxsize=2.5in
\epsfbox{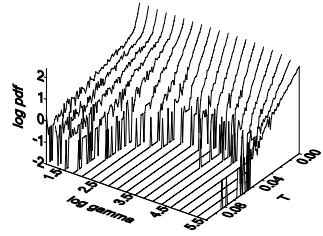}
\caption{Overall summary of the probability density 
functions of escape time measured by setting $V_{rms}=0.816$ V. 
The $z$ axis is logarithmic. From left to right the 
regimes of slow dynamics, resonant activation and fast 
dynamics are clearly seen. In the regimes of resonant 
activation and fast dynamics the probability density 
function is well described by an exponential function 
whereas in the slow dynamics regime a non-exponential 
shape emerges. \label{f.2}}
\end{figure}

\begin{figure}
\epsfxsize=2.5in
\epsfbox{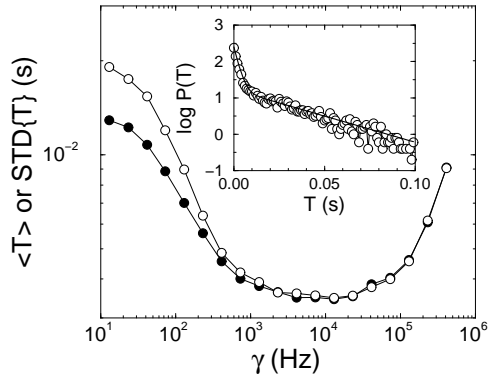}
\narrowtext
\caption{Average escape time (black circles) and its 
standard deviation (white circles) as a function of the 
correlation rate of the dichotomous noise.
The amplitude of the Gaussian noise is set 
to $V_{rms}=0.816$ V.
The difference between the two observables quantifies 
the degree of non-exponential behavior of the pdf. 
Two regimes are observed: (i) the regime of values 
of $\gamma>\mu_-$ where an exponential pdf is observed 
within experimental errors and (ii) the opposite regime
($\gamma<\mu_-$) where a non-exponential behavior 
is observed. The degree of non-exponential
behavior is maximal in the regime of slow dynamics. 
In the inset we show the probability density function 
(circles) of the escape time obtained by setting 
$\gamma=13$ Hz. The pdf is well approximated by Eq. (3) 
with $<T_+>$ and $<T_->$ measured in the absence 
of switching. This function is shown as a solid line 
in the inset.
\label{f.3}}
\end{figure}


\begin{references}

\bibitem{Kramers40} H. A. Kramers, Physica {\bf 7}, 284 (1940).

%
\bibitem{Hanggi90} P. H\"anggi, P. Talkner, M. Borkovec, Rev. Mod. Phys. 
{\bf 62}, 251 (1990).


%
\bibitem{Doering92} C. R. Doering and J. C. Gadoua, 
Phys.\ Rev.\ Lett.\ {\bf 69}, 2318 (1992). 


%
\bibitem{Zurcher93} U. Z\"urcher and C. R. Doering, 
Phys.\ Rev.\ E{bf 47}, 3862 (1993). 

%
\bibitem{Broeck93} C. Van den Broeck, Phys.\ Rev.\ 
E{\bf 47}, 4579 (1993). 

%
\bibitem{Bier93} M. Bier and R. D. Astumian, 
Phys.\ Rev.\ Lett.\ {\bf 71}, 1649 (1993);
Phys.\ Lett.\ A{\bf 247}, 385 (1998). 

%
\bibitem{Brey94} J. J. Brey and J. Casado-Pascual, 
Phys.\ Rev.\ E{\bf 50}, 116 (1994).

%
\bibitem{Pechukas94} P. Pechukas and P. H\"anggi, 
Phys.\ Rev.\ Lett.\ {\bf 73}, 2772 (1994).

%
\bibitem{Madureira95} A. J. R. Madureira {\it et al.},
Phys.\ Rev.\ E{\bf 51}, 3849 (1995).

%
\bibitem{Reimann95} P. Reimann, 
Phys.\ Rev.\ Lett.\ {\bf 74}, 4576 (1995); 
Phys.\ Rev.\ E{\bf 52}, 1579 (1995).

%
\bibitem{Iwariszewski96} I. Iwaniszewski, 
Phys.\ Rev.\ E{\bf 54}, 3173 (1996).

%
\bibitem{Reimann97} P. Reimann and T. C. Elston, 
Phys.\ Rev.\ Lett.\ {\bf 77}, 5328 (1997).

%
\bibitem{Boguna98} M. Bogu\~n\'a {\it et al.},
Phys.\ Rev.\ E{\bf 57}, 3990 (1998).

%
\bibitem{Reimann98} P. Reimann, R. Bartussek and P. H\"anggi, 
Chem.\ Phys.\ {\bf 235}, 11 (1998).  
 
%
\bibitem{Mantegna98} R.N. Mantegna and B. Spagnolo, 
J.\ Phys.\ IV {\bf 8}, 247 (1998).


%
\bibitem{Marchesoni95} F. Marchesoni and L. Gammaitoni, 
Phys.\ Lett.\ A{\bf 201}, 275 (1995).

%
\bibitem{Marchi96} M. Marchi {\it et al.},
Phys.\ Rev.\ E{\bf 54}, 3479 (1996).

%
\bibitem{Mantegna94} R.N. Mantegna and B. Spagnolo, Phys.\ Rev.\ E 
{\bf 49}, R1792 (1994); Il Nuovo Cimento D{\bf 17}, 873 (1995).

%
\bibitem{Mantegna96} R.N. Mantegna and B. Spagnolo, 
Phys.\ Rev.\ Lett.\ {\bf 76}, 563 (1996).

%
\bibitem{Bray89} A.J. Bray and A.J. McKane, 
Phys.\ Rev.\ Lett.\ {\bf 62}, 493 (1989).

%
\bibitem{Dykman90} M.I. Dykman, 
Phys.\ Rev.\ A {\bf 42}, 2020 (1990).

%
\bibitem{Hanggi85} P. H\"anggi {\it et al.},
Phys.\ Rev.\ A{\bf 32}, 695 (1985).

%
\bibitem{note1} In our test we
measure $<T>$ as a function of $1/V^2_{rms}$ for
different values of $\tau_s$. The test confirms that
our system is in a regime of colored noise because
we observe that $\Delta U_{m} \tau_s$ decreases 
when $\tau_s$ increases and approximates to a 
constant value for  $\tau_n/\tau_s < 0.01$. 


\end{references}
\end{document}